\documentclass{ws-ijgmmp}

\begin{document}

\markboth{Gabriele Gionti, S.J.}
{Some Considerations on Discrete Quantum Gravity (Paper's Title)}

%
\catchline{}{}{}{}{}
%

\title{SOME CONSIDERATIONS ON DISCRETE QUANTUM GRAVITY}

\author{GABRIELE GIONTI, S.J.}

\address{Specola Vaticana\\
Vatican City, V-00120, Vatican City State, and\\
Vatican Observatory Research Group\\
Steward Observatory, The University Of Arizona, 933 North Cherry Avenue\\
Tucson, Arizona 85721, USA\\
\email {ggionti@as.arizona.edu}}

\maketitle

\begin{history}
\received{(07 10 2011)}
\revised{(28 10 2011)}
\end{history}

\begin{abstract}
Recent results in Local Regge Calculus are confronted with Spin Foam
Formalism. Introducing Barrett-Crane Quantization in Local Regge Calculus makes 
it possible to associate a unique
Spin $j_{h}$ with an hinge $h$, fulfilling one of the requirements of Spin Foam definition. It is shown that 
inter-twiner terms of Spin Foam can follow from the closure constraint in Local Regge Calculus.

\noindent Dedicated to Beppe Marmo on the occasion of his 65th Birthday
\end{abstract}


\section{Spin Foam}	
A Spin Foam is a history, in the sense of a Feynman path integral, of spin networks (see \cite{rovelli} p. 325). 
A Spin Foam is a triple $\left( \Gamma, j_{f}, i_{e} \right)$, $\Gamma$ is a two-dimensional complex, 
$j_{f}$ are the irreducible representations of a group G associated with each face $f$ of $\Gamma$, and 
 $i_{e}$ is  an inter-twiner, which, at each edge $e$ of $\Gamma$, maps the irreducible 
representations of $G$ associated with the incoming $2$-faces at the edge $e$  
to the outcoming $G$-irreducible representations. We will restrict to the case $G=SO(4)$. Rovelli claims \cite{rovelli} p. 328, 
based on the analysis of many spin foam models, 
that the general form of the partition function for discrete quantum gravity should have the form

 \begin{equation}   
Z =\sum_{\Gamma}w(\Gamma)\sum_{J_{f},i_{e}}
\prod_{f}A_{f}(j_{f})
\prod_{e}A_{e}(j_{f},i_{e})
\prod_{v}A_{v}(j_{f},i_{e})\;\;\;\; . 
\label{foam}
\end{equation}

\noindent where $w(\Gamma)$ is a weight associated with each $2$-complex $\Gamma$, $A_{f}(J_{f})$ is a coefficient 
associated with each face $f$, and similarly $A_{e}(j_{f},i_{e})$ and $A_{v}(j_{f},i_{e})$ are coefficients associated with 
edge $e$ and vertex $v$ respectively.

\noindent There are still two unsettled problems in Spin Foam formalism: coupling with fermionic matter,  
and how Spin Foam models are related to General Relativity (i.e., the problem of the classical limit). 
A way of coupling with fermionic matter, based on Local Regge Calculus, has been addressed in \cite{ultimo}.\\
\noindent We review recent results in Local Regge Calculus in section 2 and 3 stressing some similarities between Local Regge Calculus and Spin Foam Formalism. 
Section 4 contains new results.  

\section{Regge Calculus}
\noindent An important result in Regge Calculus 
from J.Cheeger, W. Muller, R. Schrader in \cite{Cheeger} is that the Einstein-Regge action 
converges to the Einstein-Hilbert (for a review of Regge Calculus see \cite{regge}, \cite{ruth}, \cite{thesis}). 
They showed that the convergence of the action is in {\it sense of measure}.
This applies when, roughly speaking, we choose a manifold and consider an approximation of it through a triangulation 
(for example a sphere may be approximated by a triangulation inscribed in it).
The number of hinges in the simplicial manifold increases along with the number of the simplices incident at each hinge.
 The difference, in modulus, between the Einstein-Hilbert action as functional on the manifold and
the Regge-Einstein action as functional evaluated on the triangulations of the same manifold, become smaller and smaller as the triangulations will be finer and finer. \\
A different approach, with respect to the above Regge Calculus, has been proposed in \cite{Ale} with a first order formulation, in the sense of Palatini,
in \cite{Gab}. The main difference, with respect to the above Regge Calculus, consists in introducing a reference frame in
each $n$-dimensional simplex. An $SO(n)$ connection matrix $\Lambda_{\alpha\beta}$ is determined by 
two $n$-dimensional simplexes $\alpha$ and $\beta$, which share a common ($n-1$)-dimensional face $\alpha\beta$.
This matrix allows the passage from the reference frame $\alpha$ to the reference frame $\beta$.
Let $b_{\alpha\beta}(\alpha)$ be the normal vector to the common face $\alpha\beta$ in the reference frame in $\alpha$ . 
It behaves as {\it n-bein} in General Relativity \cite{frohlich}.   
Its modulo is proportional to the volume of the face itself. Let 
$b_{\beta\alpha}(\beta)$ be the same normal vector in the reference frame in $\beta$. 
These vectors are related by  

\begin{equation}
{b_{\alpha\beta}}{\;}^{a}(\alpha)=
 \Lambda^{a}_{b}(\alpha,\beta){b_{\beta\alpha}}{\;}^{b}(\beta)\;\;\;\; .
\label{relatto}
\end{equation}

\noindent (see \cite{Ale} and \cite{Gab} for details). 
In the second order formalism \cite{Ale} \cite{Gab}, the n-bein and the $SO(n)$ connection matrices are functions of the vertex-coordinates 
in the reference frame of each simplex $\alpha$. A Regge-Einstein like   
action is defined. Once one translates the formalism back into the original Regge variables, the action appears to depend on $sinK(h)$ instead of $K(h)$. 
This is generally considered a lattice artifact \cite{Ale} and  is not a problem for convergence to the continuum theory, 
since \cite{Cheeger} that convergence regime occurs when $K(h)$ is small, so that $sinK(h)\approx K(h)$. The whole theory can be considered defined on the dual-metric Voronoi-complex of
the original simplicial complex. In that dual complex, the $n$-simplexes become points, the ($n-1$)- faces dual links, and the $n-2$-hinges become two-dimesional plaquettes \cite{Ale} \cite{Gab}.      
 Labeling the vertices of the dual plaquette $h$ $\left\{\alpha,\beta,\delta^{h}_{1},...,\delta^{h}_{k}\right\}$ , as in Lattice Gauge Theory, the holonomy around the plaquette is defined as 

\begin{equation}
U^{h}_{\alpha \alpha}\equiv
\Lambda_{\alpha\beta}\Lambda_{\beta\delta^{h}_{1}}
...\Lambda_{\delta^{h}_{k}\alpha}\;\;\;\;,
\label{holonomy}
\end{equation}

\noindent where $\Lambda_{\alpha\beta},\;\Lambda_{\beta\delta^{h}_{1}},\;...\Lambda_{\delta^{h}_{k}\alpha}$ are the matrices associated, respectively, with the links 
$\alpha\beta,\;\beta \delta^{h}_{1},...,\;\delta^{h}_{k}\alpha$ of the plaquette $h$. The bivector \cite{Ale} \cite{Gab}, in the reference frame $\alpha$, whose modulus is equal to the volume of the $n-2$-dimensional hinge, is defined as 

 \begin{equation}
{\mathcal{V}}^{(h)}{\;}^{c_1c_2}(\alpha)\equiv
{1\over n!(n-2)!V(\alpha)}\Bigl({b_{\alpha\beta}}^{c_1}
(\alpha){b_{\alpha\delta^{h}_{k}}}^{c_2}(\alpha)
-{b_{\alpha\beta}}^{c_2}(\alpha){b_{\alpha\delta^{h}_{k}}}^{c_1}(\alpha)\Bigr),
\label{arrota}
\end{equation}

\noindent where $V(\alpha)$ is the volume of the $n$-dimensional simplex. However, this is still a second-order formalism of discrete General Relativity.
Now we introduce a first-order formalism in which $\Lambda_{\alpha\beta}$ and
$b_{\alpha\beta}(\alpha)$ are independent variables. Then equation (\ref{relatto}) becomes a constraint

\noindent  There is a further constraint on the $b_{\alpha\beta}(\alpha)$, 
since  the $n+1$ normals to the ($n-1$)-dimensional faces
 of a $n$-simplex $\alpha$ are linearly dependent,

\begin{equation}
\sum_{\beta=1}^{n+1}b_{\alpha\beta}(\alpha)=0
\label{chiusura}\;\;\;\; .
\end{equation}

 \noindent In order to define a gravitational action
independent of the starting simplex reference frame,
 we need to define the following
antisymmetric tensor on the plaquette $h$

\begin{eqnarray}
W^{(h)}_{c_1c_2}(\alpha) &\equiv &
{1\over k_{h}+2} \Big({\mathcal{V}}^{(h)}(\alpha)+
\Lambda_{\alpha\beta}{\mathcal{V}}^{(h)}(\beta)\Lambda_{\beta\alpha}+... \nonumber\\
&+&\Lambda_{\alpha\beta}...
\Lambda_{\delta^{h}_{k-1}\delta^{h}_{k}}
{\mathcal{V}}^{(h)}(\delta^{h}_{k})\Lambda_{\delta^{h}_{k}\delta^{h}_{k-1}}...
\Lambda_{\beta\alpha} \Big)_{c_1c_2}
\label{media}\;\;\;\;.
\end{eqnarray}

\noindent The action then is

\begin{equation}
S\equiv -{1\over 2}\sum_{h}{\rm Tr}
\left(U^{h}_{\alpha\alpha}W^{h}(\alpha)\right)
\label{caction}
\end{equation}

\noindent The action (\ref{caction}) coincides with the action in the second-order formalism \cite{Ale} \cite{Gab}.
From this point on, we shall focus mainly on the four
dimensional case. The action (\ref{caction}) is invariant under the $SO(4)$ group, which
is the gauge group. It has been shown that the discrete equations of motion for this theory 
can be derived, \cite{thesis} \cite{Gab} , and appear very similar to the General Relativity equations in the covariant tetrad formalism \cite{myself}.
In the limit of {\it small deficit angles}, the {\it Levi-Civita-Regge} connection \cite{Gab} is, locally, the unique solution. 
Clearly this is in total correspondence with the Palatini continuum theory in General Relativity 
where, in the torsion-free case, the Levi-Civita connection is solution of the field equations for the connection.

\noindent This first order Regge Calculus is particularly well suited for coupling between gravity and fermionic matter. 
As has been pointed out in \cite{Gab}, a Dirac spinor $\psi(\alpha)$ can be defined at each vertex $\alpha$ of the dual-Voronoi 
complex. A two-fold covering matrix group $D_{\alpha\beta}$ of $SO(4)$ replaces the connection matrices $\Lambda_{\alpha\beta}$. 
The total action is the sum of the Regge-Einstein action and discrete Dirac action.   

\section{Heading toward Spin Foam}
\noindent In the following we will use the notation: $\mu(b_{\alpha\beta}(\alpha))\equiv
db^{1}_{\alpha\beta}(\alpha)...db^{5}_{\alpha\beta}(\alpha)$, and $ \mu(\Lambda_{\alpha\beta})$ for the Haar measure on $SO(4)$. 
Taking into account all the constraints that this theory carries \cite{Gab}, the path integral associated with the Local 
Regge-Einstein equation (\ref{caction}) of the previous section, 
is
 \begin{equation}
Z=\int e^{{1\over 2}\sum_{h}Tr\left(U^{h}_{\alpha\alpha}W^{h}(\alpha)\right)}
\prod_{\alpha}\delta(\sum_{\beta=1}^{5}b_{\alpha\beta}(\alpha))
\prod_{\alpha\beta}
\delta\left(b_{\alpha\beta}(\alpha)-\Lambda_{\alpha\beta}b_{\beta\alpha}(\beta)\right)
\mu(\Lambda_{\alpha\beta})\mu(b_{\alpha\beta})\;\;\;\;.
\label{part}
\end{equation}

\noindent Some observations from Group Theory (see \cite{ultimo} for all details) suggest the expansion of the exponential of the action
in characters $\chi_{J}\left(U^{h}\right)$ of the irreducible
representations $J$ of the group $SO(4)$, as in lattice Gauge Theory. We can omit the index $\alpha\alpha$ in the holonomy
matrix $U^{h}$, since it does not depend on the starting simplex $\alpha$ \cite{Gab}, 

\begin{equation}
e^{{1\over 2}Tr\left(U^{h}_{\alpha\alpha}W^{h}(\alpha)\right)}=\sum_{J}
c_{J}^{h}\Big(b_{\alpha\beta},b_{\beta\alpha},...,
b_{\delta^{h}_{k}\alpha},b_{\alpha\delta^{h}_{k}}\Big)\chi_{J}\left(U^{h}\right)\;\;\;\;,
\label{svilup}
\end{equation}

\noindent where the coefficients $c_{J}^{h}\Big(b_{\alpha\beta},b_{\beta\alpha},...,
b_{\delta^{h}_{k}\alpha},b_{\alpha\delta^{h}_{k}}\Big)$ are defined by 

\begin{equation}
c_{J}^{h}\Big(b_{\alpha\beta},b_{\beta\alpha},...,b_{\delta^{h}_{k}\alpha},b_{\alpha\delta^{h}_{k}}\Big)\equiv\int_{SO(4)}\mu\big(U^{h}\big)e^{{1\over
2}Tr\left(U^{h}_{\alpha\alpha}W^{h}(\alpha)\right)}
{\overline{\chi_{J}\left(U^{h}\right)}}\;\;\;\;.
\label{coeffio}
\end{equation}

\noindent Therefore the partition function for discrete quantum gravity becomes

\begin{eqnarray}
Z=\int \prod_{h}\sum_{J}c_{J}^{h}\Big(b_{\alpha\beta},b_{\beta\alpha},...,
b_{\delta^{h}_{k}\alpha},b_{\alpha\delta^{h}_{k}}\Big)\chi_{J}\left(U^{h}\right)
\prod_{\alpha}\delta\big(\sum_{\beta=1}^{5}b_{\alpha\beta}(\alpha)\big)\nonumber\\
\prod_{\alpha\beta}
\delta\big(b_{\alpha\beta}(\alpha)-\Lambda_{\alpha\beta}b_{\beta\alpha}(\beta)\big)
\mu(\Lambda_{\alpha\beta})\mu(b_{\alpha\beta})\;\;\;\;.
\label{color}
\end{eqnarray}

\noindent The equations above highlight the fact that each two-dimensional plaquette is associated with irreducible representations $J$ of 
$SO(4)$. This association is not unique, as is clear from the coefficients of the characters $\chi_{J}\left(U^{h}\right)$ expansion 
in (\ref{svilup}), there is a numerable set of $SO(4)$ irreducible representations $J$ associated with each plaquette $h$.  
 A further step is needed so that an irreducible representation $J_{h}$ of $SO(4)$ will be uniquely associated with the  plaquette $h$, as in Spin Foam Formalism. 
The partition function (\ref{color}) resembles a Spin Foam partition function with respect to the products on the plaquettes, the links, and the vertices (see eq. \ref{foam} sec.1).

\section{Barrett-Crane Quantization}

\noindent Barrett and Crane \cite{B-C} propose a model for deriving spin networks by considering the double covering of $SO(4)$, $SU(2)\times SU(2)$. 
In this approach, Relativistic quantum spins of $SU(2)\times SU(2)$ are associated with the two-dimensional hinges of the simplicial complex. 
We apply these ideas to Local Regge Calculus in order to find new results.
The self-dual and anti-self-dual parts of each bi-vector (eq. \ref{arrota}, sec. 2) 
form a representation of the $SU(2)\times SU(2)$ Lie algebra $su(2) \oplus su(2)$ \cite{B-C}. Since the bi-vectors are simple (i.e, they are wedge product of vectors),
they correspond to $(j,j)$ representations of $SU(2)\times SU(2)$ \cite{B-C}. The next step is quantization, which Barrett and Crane implement with a map from 
vectors on $\Re^{4}$ to vector operators on Hilbert Space \cite{B-C}. In our case, we get a map from the dual and anti-self-dual components of each bi-vector to 
the corresponding two sets  of Angular Momentum
Operatrors $( {\bf{J}^{+}}^{2}, {\bf{J}^{+}}_{x}, {\bf{J}^{+}}_{y},{\bf{J}^{+}}_{z} )$ and  
$( {\bf{J}^{-}}^{2}, {\bf{J}^{-}}_{x}, {\bf{J}^{-}}_{y},{\bf{J}^{-}}_{z} )$ \cite{Rubvelli}. 

\noindent In \cite{Rubvelli}, it is shown that the square modulus of each bi-vector is twice the square modulus of either its self-dual or anti-self-dual part. 
Rovelli (\cite{rovelli} p.249) shows that, for SU(2)  spin networks, the area  is quantized. Since the self-dual and anti-self-dual parts of each 
bi-vector component are independent (i.e, they form a direct sum) and since each one is independently  an $SU(2)$ representation, 
we can finally deduce that the square moduli of the self-dual and anti-self-dual components of each bi-vector is proportional resepctively  
to the ${\bf{J}^{+}}^{2}$ and ${\bf{J}^{-}}^{2}$  eigenvalues, as for  SU(2) spin networks. 
These factors indicate that  the quantum area $V(h)$ of a hinge $h$, whose associated $SO(4)$ 
spin representation  is $(j_{h},j_{h})$, is  $V(h)=8\pi G c^{-3}\hbar  \sqrt{2j_{h}( j_{h}+1)}$, in the correct physical units. In the second-order  
formalism, the case in which the connection matrices $\Lambda_{\alpha\beta}$ depend on the vertex coordinates, the local Regge-Einstein action
is $S=\sum_{h} sin\theta (h) V(h)$ \cite{Ale}, $\theta(h)$ being the deficit angle at the hinge $h$ \cite{regge}. Implementing Barrett-Crane quantization, the action becomes 
$S=\sum_{h}8\pi \hbar G c^{-3}  sin\theta(h) \sqrt{2j_{h}(j_{h}+1)}$, which is still dependent on the $n-bein$ vectors 
$b_{\alpha\beta}(\alpha)$, since the deficit angle $\theta(h)$ is linear in the dihedral angles among the $n-1$-faces which share the hinge $h$.
The dihedral angles can be written in terms of the normals $b$ to the $n-1$ faces \cite{regge}.

\noindent Therefore the partition function is

\begin{eqnarray}
Z=\sum_{J_{\sigma}} \int \prod_{h}e^{-\left(8\pi \hbar G c^{-3}sin\theta(h)\sqrt{2j_{h}(j_{h}+1)} \right)}
\prod_{\alpha}\delta(\sum_{\beta=1}^{5}b_{\alpha\beta}(\alpha))\nonumber\\
\prod_{\alpha\beta}
\delta\left(b_{\alpha\beta}(\alpha)-\Lambda_{\alpha\beta}b_{\beta\alpha}(\beta)\right)
\mu(b_{\alpha\beta}(\alpha))\;\;\;\;.
\label{seconquan}
\end{eqnarray}

\noindent In the second-order formalism, the matrices  $\Lambda_{\alpha\beta}$ can be expressed as functions of 
the $b_{\alpha\beta}$ \cite{Ale}. Notice that the Barrett-Crane quantization makes it possible to associate a unique irreducible representation $j_{h}$   \
 with each hinge $h$. 
This is one of the requirements in the definition of Spin Foam. Without quantizing
(see section 3 eq.\ref{color}) there is no unique way to make this correspondence. 
The sum over $J_{\sigma}$ accounts for inequivalent ways of associating an irreducible representation $j_{h}$ of $SU(2)$ to each hinge $h$. 
This is the simplest  proposal for Barrett-Crane Spin Foam amplitude which has been derived from 
Local-Regge Calculus in the second-order formalism. 
As anticipated in section 1 eq.\ref{foam}, the partition function can be expressed as the product of three factors: 
a product on the  hinges, a product on the links, and a product on the vertices.  
Contrary to what is usually claimed in standard Spin Foam Formalism, inter-twiner coefficients are completely absent. 
In the usual approach to Spin Foam Formalism \cite{rovelli} p. 333, the partition function has an imaginary
exponential. This choice comes from the asymptotic behavior of  
the Ponzano-Regge state sum model in three dimensional discrete quantum gravity \cite{rovelli} p.335-36, in the limit of large angular momentum.
This imaginary exponential implies that the integration over the "metric" variable (i.e., a edge length in three dimensions) 
\cite{rovelli} p.333 generates a delta function, 
which can be expanded as a sum over the traces of the unitary irreducible representations of the group. 
Then integration over the group variables becomes a product of inter-twiners. But it is not our case. Wick rotation  implies that the partition function exponential 
is the negative of the action, which is a real term in the Local Regge Calculus case. Therefore the integration over the "metric" variables does not become a delta-function, 
and the inter-twiner coefficients are absent in the path integral.

\noindent A natural question to rise at this point is whether we can derive something like inter-twiner coefficients of this 
Spin Foam amplitude. 
Interestingly enough, the closure constraint on the $n-bein$ $b$, section 2; eq. \ref{chiusura}, 
implies that if we fix the link $\alpha\beta$ and the corresponding $b_{\alpha\beta}^{a}(\alpha)$ 
and sum over all other $n-beins$ originating from the dual vetrex $\alpha$, we get 

\begin{equation}
\sum_{\delta^{h}_{k}\neq \beta} {\mathcal{V}}^{(h)}{\;}^{c_1c_2}(\alpha)=
\sum_{\delta^{h}_{k}\neq \beta}{1\over n!(n-2)!V(\alpha)}\Bigl({b_{\alpha\beta}}^{c_1}
(\alpha){b_{\alpha\delta^{h}_{k}}}^{c_2}(\alpha)
-{b_{\alpha\beta}}^{c_2}(\alpha){b_{\alpha\delta^{h}_{k}}}^{c_1}(\alpha)\Bigr)=0.
\label{relatto1}
\end{equation}

\noindent In four dimensions, four plaquettes (i.e, metric duals of the corresponding hinges) share the link $\alpha\beta$
 on the same dual vertex $\alpha$. The corresponding four bi-vectors are linearly dependent through equation \ref{relatto1}. 
Therefore, the irreducible $SO(4)$ representations,$J_{1}, J_{2}, J_{3}, J_{4}$, associated with these bi-vectors are linear dependent 

\begin{equation}
J_{1}+J_{2}+J_{3}+J_{4}=0.
\label{depe}
\end{equation}

\noindent These considerations hold for each (dual) link in this Spin Foam Formalism, 
and generate the BC (Barrett and Crane)-inter-twiners (see \cite{rovelli} p. 348-356). 
Finally, we can conclude that the inter-twiner coefficients correspond to the closure constraint.

\section{Conclusions}
We have continued to search for closer relations between Local Regge Calculus and Spin Foam formalism. One of the main reasons for this
research is that Local Regge Calculus converges, in measure, to continuum General Relativity, and it can be easily coupled with fermionic matter. 
Therefore any bridge between these two approaches to Discrete Quantum Gravity would help to throw light on both the convergence of Spin Foam Formalism to 
General Relativity and coupling with matter.\\ 
\noindent We started, in section 2, with a very brief review of the basic concepts 
of Local Regge Calculus in the first order formalism. Section 3 reviewed the path integral expansion of Local Regge Calculus in group-characters, as in Lattice Gauge Theory. 
The form of this expansion showed a resemblance with Spin Foam amplitude. Following the main ideas of Barrett and Crane,  a new further quantization was proposed in section 4. 
This quantization allowed a unique and un-ambiguous correspondence between each irreducible representation $j_{h}$  and 
the hinge $h$. It also resulted in a quantization of hinge areas.
Using Local Regge Calculus, we also proposed an amplitude for Spin Foam. By contrast, a Spin Foam amplitude based on Local Regge Calculus  does not have inter-twiner terms. 
We noted that inter-twiner terms can be seen as a derivations of the closure constraints of Local Regge Calculus.

\end{document}